\documentclass[prr,twocolumn,groupedaddress,amsmath,amssymb,amsfonts]{revtex4-2}
\usepackage{txfonts}
\usepackage{bm} 
\usepackage{algorithm2e}
\usepackage{subfigure}
\usepackage{mathrsfs}
\usepackage{array}
\usepackage{amsmath,latexsym}
\usepackage{amsfonts} 
\usepackage[dvipsnames]{xcolor}
\usepackage{amssymb} 
\usepackage[margin=3cm]{geometry}
\usepackage{braket}
\usepackage{comment} 
\usepackage{graphicx, color}
\usepackage[colorlinks=true, bookmarks=true,
bookmarksnumbered=true, bookmarkstype=toc, linkcolor=RoyalBlue,
urlcolor=CadetBlue, citecolor=RoyalBlue]{hyperref}
\newcommand{\ST}[1]{\textcolor{blue}{#1}}

\begin{document}
\title{Deep Unfolded Local Quantum Annealing
}
\author{Shunta Arai}
\email[]{arai.s.ba03@m.isct.ac.jp}
\author{Satoshi Takabe }
\affiliation{Institute of Science Tokyo, Ookayama, Tokyo 152-8550, Japan}
\date{\today}
\begin{abstract}
Local quantum annealing (LQA), an iterative algorithm, is designed to solve combinatorial optimization problems.
It draws inspiration from QA, which utilizes adiabatic time evolution to determine the global minimum of a given objective function.
In the original LQA, the classical Hamiltonian is minimized via gradient descent. 
The performance of LQA depends on the choice of the parameters. 
Owing to the non-convex nature of the original cost function, LQA often becomes trapped in local minima, limiting its effectiveness.
To address this challenge, we combine LQA with a deep unfolding scheme, which enables us to tune the parameters from the training data via back-propagation. 
{As a demonstration, we apply the deep unfolded LQA to the Sherrington-Kirkpatrick model, which is a fundamental {model} in statistical physics.}
Our findings exhibit that deep unfolded LQA outperforms the original LQA, exhibiting remarkable convergence speed and performance improvement.
As the trained parameters can be generalized to unknown instances and different system sizes, our results have significant practical implications and provide valuable insights for real-world applications. 
\end{abstract}
\maketitle
\section{Introduction}
\label{sec1}
Metaheuristics, such as simulated annealing (SA) \cite{Kirkpatrick_1983} and quantum annealing (QA) \cite{Kadowaki_1998,Farhi_2001,santoro_2006,Das_2008,Albash_2018,Grant2020,Crosson_2021}, are powerful techniques for solving combinatorial optimization problems. 
These algorithms, rooted in statistical and quantum physics, have been widely adopted because they efficiently obtain near-optimal solutions to complex problems.
{Recent advancements have led to the development of specialized hardware using annealing algorithms, known as \textit{Ising machines}, for example, the D-Wave machine \cite{Dwave2010a, Dwave2010b, Dwave2010c, Dickson_2013}, Fujitsu's digital annealer \cite{Aramon_2019}, and the simulated bifurcation machine \cite{Goto_2021}. 
They are mainly designed for solving quadratic unconstrained binary optimization (QUBO) problems \cite{Glover_2022} that contain a broad class of combinatorial optimization problems. 
Benchmark studies have been performed to explore the potential of the Ising machine \cite{oshiyama_2022,Kowalsky_2022,Mohseni_2022,Jiang_2023}.
By utilizing QA or advanced computational techniques, such as GPU and high-performance computation, these machines are expected to meet the growing demand for {obtaining better solutions faster}, outperforming traditional SA on classical computers.}

As an alternative approach to solving the QUBO using a classical computer, \textit{local QA} (LQA) has been proposed \cite{Bowles_2022}.
LQA is based on the semiclassical approximation of the Hamiltonian of QA, which considers the product state and solves the semiclassical Hamiltonian's minimization problem using gradient descent (GD). 
{Since all variables are updated parallelly at each step, LQA can be applied with parallel computing like GPU, leading to high computational efficiency \cite{Bowles_2022}.}
The performance of LQA depends on the annealing schedule and parameters of GD.
Appropriate parameters lead to a reduction in the execution time of LQA and an enhancement of its performance.
A similar situation can be observed in QA and the quantum approximate optimization algorithm (QAOA) \cite{farhi_2014, Brady_2021}. 
Although appropriate parameter settings are essential, the optimization problem for determining hyperparameters is NP-hard \cite{Bittel_2021}. 
Heuristics are required to optimize the parameters.

{
In the context of iterative optimization algorithms for signal processing, \textit{deep unfolding} {(DU)} or \textit{unrolling} were developed to tune {parameters of an algorithm} with the aid of techniques developed in deep learning \cite{Gregor_2010, Balatsoukas_2019}.
In the DU scheme, {an iterative algorithm is} unfolded into a sequence of procedures {as}  layers. 
We embed trainable parameters into each layer, {which controls the performance of the algorithm}. 
The unfolded network can be interpreted as a feed-forward neural network.
If each process is differentiable, the trainable parameters are learned by minimizing {a} loss function and updated via back-propagation \cite{Shlezinger_2022}.
The data-driven tuning of parameters yields performance improvement for the iterative algorithms and accelerates its convergence \cite{takabe_2021}. }

In a {previous} study, {for instance,} the DU scheme was applied to solve a multiple-input multiple-output (MIMO) problem \cite{Larsson_2014}, which was formulated as a combinatorial optimization problem.
Representative examples of DU-based MIMO detectors include the Habburd--Stratonovish detector \cite{takabe_2024} and the simulated bifurcation \cite{takabe_2023}.
The DU scheme has also been utilized to accelerate the Monte Carlo-based solver for QUBO with linear constraints \cite{Hagiwara_2024}.

 In this study, we apply the DU scheme to LQA to improve its performance and accelerate its convergence speed.
 We refer to LQA with the DU scheme as \textit{deep unfolded LQA} (DULQA). 
 We compare the performance of DULQA to that of LQA using GD and Adam \cite{kingma_2017}. 
 {As {a} benchmark of DULQA, we solve the Sherrington-Kirkpatrick model \cite{Sherrington_1975}, which is familiar {in} the statistical physics  \cite{nishimori_2001}.}
 We demonstrate that DULQA outperforms LQA by reaching lower energy states faster.
 This performance improvement is consistent across unknown instances and system sizes. 
 In addition, we perform finite-size scaling of the attained observables and analyze the scaling exponents for different learning strategies.
 This study provides a widely applicable DU-based optimization algorithm for combinatorial optimization problems.

 The remainder of this paper is organized as follows:
{ In Sec.\ref{sec2}, we explain the mathematical formulation of LQA and the principle of the DU scheme.
 In Sec.\ref{sec3}, we introduce the structure and training protocol of DULQA.}
 In Sec.\ref{sec4}, we describe the problem and parameter settings in our simulations and demonstrate the numerical results.
 In Sec.\ref{sec5}, we conclude the study and discuss the future research directions.
 
\section{Preliminaries}
\label{sec2}
{This section briefly shows the mathematical definition of LQA and the concept of the DU scheme. 
First, we review the vanilla LQA.
Second, we provide an editorial review of the DU scheme.}
\subsection{Local quantum annealing}\label{sec21}

To formulate LQA, we begin with the standard Hamiltonian of QA as follows:
\begin{align}
    \hat{H}(s(t),\gamma)& = s(t) \gamma \hat{H}_0 - (1 - s(t))\hat{H}_{\mathrm{TF}},\label{eq1}\\
    \hat{H}_0& = \sum_{i\neq j} J_{ij} \hat{\sigma}_i^z\hat{\sigma}_j^z + \sum_{i = 1}^Nh_i\hat{\sigma}_i^z,\label{eq2}\\
   \hat{H}_{\mathrm{TF}}& = \sum_{i = 1}^N\hat{\sigma}_i^x \label{eq3},
\end{align}
where $\hat{\sigma}_i^{z}$ and $\hat{\sigma}_i^{x}$ denote the Pauli operators acting on site $i$, $h_i$ and $J_{ij}$are the parameters of the problem to be solved, $N$ is the number of spins, and $\gamma$ represents the control parameter of the relative strength in Eq. \eqref{eq2}. The annealing parameter $s(t) = t / \tau\in [0,1]$ controls the energy contributions of the two Hamiltonians.
 In vanilla QA, the system starts from the ground state in Eq. \eqref{eq3} and evolves from $t = 0$ to $t = \tau$ following the Schr\"odinger dynamics. 
 If the state evolution is adiabatic, the system maintains an instantaneous ground state, and we can obtain the ground state of Eq. \eqref{eq2} at $t = \tau$.

\begin{figure*}[t]
\centering
\includegraphics[width=155mm]{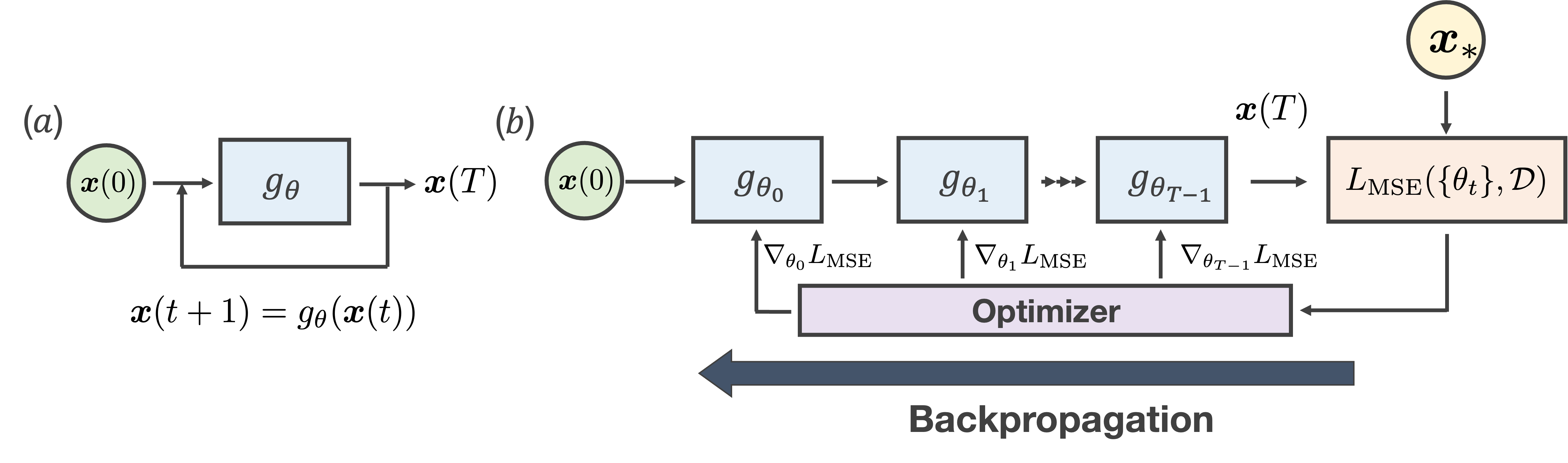}
\caption{(a): Schematic diagram of an iterative algorithm with $g_\theta$.
(b): DU-based training protocol for supervised learning. }
\label{fig:fig1}
\end{figure*}

The total Hilbert space in Eq. \eqref{eq1} exponentially increases as a function of $N$. 
To avoid treating the total Hilbert space, {we apply the semiclassical approximation of Eq. \eqref{eq1} with the spin-1/2 coherent state }\cite{Muthukrishnan_2016} characterized by $\phi_i \in[0,2\pi )$ as follows: 
\begin{align}
\ket{\bm{\phi}} = \otimes_{i = 1}^N\left[\cos \frac{\phi_i}{2}\ket{+} + \sin \frac{\phi_i}{2}\ket{-}\right]\label{eq4},
\end{align}
where $\ket{+} = (\ket{0} + \ket{1})/\sqrt{2}$ and $\ket{-} = (\ket{0} - \ket{1})/\sqrt{2}$.
{Using Eq. \eqref{eq4}, the total Hilbert spaces are restricted in the product states, and then the classical Hamiltonian is given by :} 
\begin{align}
&C(\bm{\phi},\bm{h},\bm{J},s(t),\gamma)\nonumber\\
& = \bra{\bm{\phi}}\hat{H}(s(t),\gamma)\ket{\bm{\phi}}\nonumber\\
& = s(t)\gamma \left(\sum_{i \neq j}J_{ij}\sin\phi_i\sin\phi_j + \sum_{i = 1}^Nh_i\sin\phi_i\right) \nonumber\\
&–(1 - s(t))\sum_{i = 1}^N \cos \phi_i,
\label{eq5}
\end{align}
where $\bm{h}$ and $\bm{J}$ denote the vector and matrix representations of  $h_i$ and $J_{ij}$, respectively.

Based on a previous study \cite{Bowles_2022}, we transform $\phi_i$ into {$\phi_i = \pi \tanh(w_i)/2$}
with $w_i\in \mathbb{R}$.
In this transformation, the classical Hamiltonian is represented as follows: 
\begin{align}
&C(\bm{w},\bm{h},\bm{J},s(t),\gamma)\nonumber\\
& = s(t)\gamma \left(\bm{z}^T\bm{J}\bm{z} + \bm{h}^T\bm{z}\right) - (1 - s(t))\bm{1}^T\bm{x}
\nonumber\\
&\equiv E_w, \label{eq6}\\
\bm{z}& = \left[\sin\left(\frac{\pi}{2}\tanh w_1\right),\dots,\sin\left(\frac{\pi}{2}\tanh w_N\right)\right]^T,
\label{eq7}\\
\bm{x}& = \left[\cos\left(\frac{\pi}{2}\tanh w_1\right),\dots,\cos\left(\frac{\pi}{2}\tanh w_N\right)\right]^T,
\label{eq8}
\end{align}
where $\bm{1}^T$ is defined as $\bm{1} = (1,\dots,1)$. 
We adopt GD to minimize Eq. \eqref{eq6}.
We initialize the relaxed spin parameters as $w_i(t = 0)\sim (2u_i - 1)f$, where $u_i$ is a random number sampled from $Unif[0,1)$ and $f$ is {the scale parameter of $w_i(t = 0)$.}
{Small $f \ll1 $ corresponds to  $\cos \phi_i\simeq 1$. 
The spin state is oriented to the $\ket{+}$ direction. 
On the other hand, large $f\gg1$ represents $\sin \phi_i \simeq \pm1$. 
The corresponding spin state is the $\ket{0}$ or $\ket{1}$ direction.}
 
The update rule for $\bm{w}(t)$ is as follows:
 \begin{align}
    \bm{w}(t + 1)& = \bm{w}(t) –\eta \nabla_{\bm{w}(t)} C(\bm{w}(t),\bm{h},\bm{J},s(t),\gamma)\label{eq9},
\end{align}
where $\eta$ is the step size.
At each time step $t$, the classical spin configuration is recovered by $\bm{\sigma} = \mathrm{sgn}(\bm{w}(t))\in\{\pm1\}^N$. 
The energy of the Ising spin configuration is defined as $E_{\mathrm{ising}}\equiv \sum_{i\neq j} J_{ij} \sigma_i\sigma_j+\sum_{i = 1}^Nh_i\sigma_i$.
The performance of this algorithm depends on the initial states $\bm{w}(t = 0)$ and parameters $\eta$ and $\gamma$. The selection of parameters is a crucial problem.

{
\subsection{Deep unfolding}\label{sec22}
In the DU scheme, we introduce time dependence in the parameters of {an} iterative algorithm to realize flexible algorithm control.
{These trainable parameters are learned to reflect on the statistical properties of problem instances via end-to-end {learning}.}
In this subsection, we consider supervised learning to simplify the explanation of the DU scheme. 
In supervised learning, the training dataset is given by $\mathcal{D}=\{(\bm{x}^{d},\bm{x}_*^{d})\}_{d=1}^{|\mathcal{D}|} $ where $\bm{x}^{d}\in \mathbb{R}^{N}$ is the initial position of {a} vector, $\bm{x}_*^{d}$ is the target solution of the iterative algorithm and $|\mathcal{D}|$ represents the size of training dataset. 
The general update equation of the iterative algorithm is given by $\bm{x}(t+1)=g_{\theta_t}(\bm{x}(t))$, where $g_{\theta_t}: \mathbb{R}^N\rightarrow \mathbb{R}^N$ is the update function, $\bm{x}(t)$ is the position vector at the $t$-th iteration and $\theta_t$ are the time-dependent trainable parameters.
\ST{A} simple example of $g_{\theta_t}(\bm{x}(t))$ is the gradient descent as $g_{\theta_t}(\bm{x}(t))=\bm{x}(t)-\eta(t)\nabla_{\bm{x}}f(\bm{x}(t))$ where $f(\bm{x})$ is the target function to be minimized and $\theta_t=\eta(t)$ is the step-size.
We iterate the update equation $T$ times.
In iterative algorithms, the current output $\bm{x}(t+1)$ is the next input of the update equation.
In the DU scheme, we unfold this iterative process in the time direction and regard it as a feed-forward neural network shown in Fig. \ref{fig:fig1}
}

{
As used in deep learning, we {attempt to minimize a} loss function between the final output of the update equation and the solution. {An example of a loss function is the mean squared loss defined by}
\begin{align}
    L_{\mathrm{MSE}}(\{\theta_t\},\mathcal{D})=\frac{1}{|\mathcal{D}|}\sum_{d=1}^{|\mathcal{D}|}||\bm{x}^d(T)-\bm{x}_*^d ||_2^2.
\end{align}
If the update function is differentiable, the time-dependent trainable parameters $\{\theta_t\}_{t=0}^{T-1}$ can be learned from the training dataset via back-propagation.  
The time-dependent flexible control of parameters accelerates the convergence of iterative algorithms and enhances its performance in {many} cases \cite{takabe_2021}.
A more detailed explanation of the DU scheme and its application are studied in these reviews \cite{Balatsoukas_2019, Monga_2021}
}
\section{Deep unfolded local quantum annealing}\label{sec3}

We focus on the learning of $\eta$ and $\gamma$ in LQA \footnote{{
In the original LQA \cite{Bowles_2022}, $\eta$ and $\gamma$  are introduced as hyperparameters. 
Throughout the paper, these hyperparameters are simply referred to as parameters.}}.
In the DU scheme, the parameters to be trained depend on $t$ as $\theta_t = (\eta(t),\gamma(t))$ to enhance the flexibility of the algorithm. 
This is reasonable because the step size of GD $\eta(t)$ and the control parameter $\gamma(t)$ should change depending on the energy landscape of each $t$ during the annealing process.
Therefore, the proposed DULQA with an annealing time of $\tau$ has {$2(\tau+1)$} trainable parameters $\Theta_\tau = \{\theta_t\}_{t=0}^\tau$.
{As explained in Sec.~\ref{sec22}, we can} construct a feed-forward network by borrowing the structure of LQA, which has a relatively small number of trainable parameters.
{Since {target solutions} required for supervised learning are hard to estimate in advance, we {alternatively} introduce unsupervised learning.}

{{In unsupervised learning of DULQA,}} the loss function is defined as follows: 
\begin{align}L(\Theta_t;\mathcal{D}) = \frac{1}{|\mathcal{D}|}\sum_{d = 1}^{|\mathcal{D}|}C(\bm{w}^d(\tau + 1),\bm{h}^d,\bm{J}^d,1,\gamma(\tau)),
\label{eq10}
\end{align}
where {the training data} is {defined by} $\mathcal{D} = \{\bm{w}^{d}(0),\bm{h}^d,\bm{J}^d\}_{d = 1}^{|\mathcal{D}|}$.
{{Since the training data do not contain target solutions, the training process is performed unsupervised} as in the previous study \cite{Hagiwara_2024}.} 
The trainable parameters $\Theta_t$ are updated using a GD-based optimizer, where the gradient for $\Theta_t$ is estimated by backpropagation.

\begin{figure}[t]
\centering
\includegraphics[width=70mm]{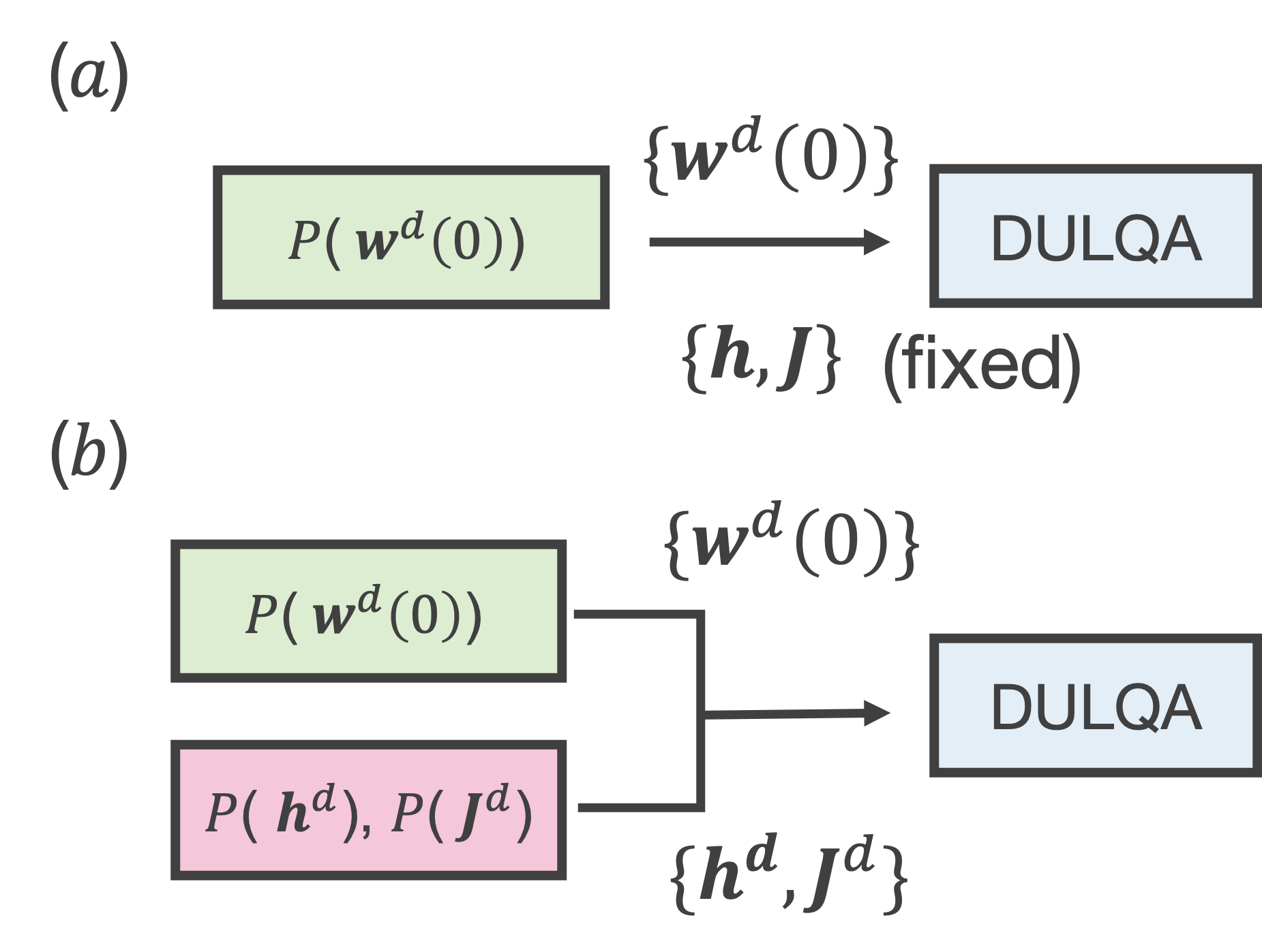}
\caption{The learning strategies in DULQA (a): 1-instance learning and (b): ensemble learning. }
\label{fig:fig2}
\end{figure}

Learning methods can be divided into two strategies depending on the generation of a dataset.
The first method is \textit{ensemble learning}, where new datasets are generated for each epoch. 
The datasets used in the previous epoch differ from those used in the current epoch.
{The ensemble learning was applied in the DU-based MIMO detector \cite{takabe_2019a,takabe_2019b}.}
{As the second method, we introduce \textit{1-instance learning},} where 
problem $\{\bm{h},\bm{J}\}$ is fixed during the {training} process, and {an initial point $\bm{w}^d(0)$ changes randomly} in each epoch.
{The advantage of 1-instance learning is that it reduces the computational cost of generating a training dataset, and it can be applied when the amount of data is limited, whereas ensemble learning can adopt the fluctuation of randomly generated problems.
The schematic diagram of two learning strategies is summarized in Fig. \ref{fig:fig2}.}

{
Following the standard training procedure of DU,} we employ incremental training \cite{Takabe_2019}.
Considering a DULQA with a large $\tau$,
a vanishing gradient occurs \cite{Hochreiter_1998}.
This hinders the deep structure of the algorithm from learning and leads to poor performance.
Incremental training aims to prevent a vanishing gradient and enhance learning accuracy.
In incremental training, we begin by learning hyperparameters $\Theta_0=\theta_0$ with the first update in Eq. (\ref{eq9}). Next, $\Theta_1$ are learned, with the initial values of $\Theta_0$ set to the previously learned ones. 
The training process is repeated until $t$ reaches $\tau$ and all $\Theta_\tau$ values are learned.
We summarize the detailed learning protocol of DULQA in Alg. \ref{alg:1}.
The outer loop in Alg. \ref{alg:1} corresponds to the incremental training.

\RestyleAlgo{ruled}
\begin{algorithm}[t]
\caption{Learning protocol of DULQA with a linear annealing schedule }\label{alg:1}
 \SetKwInOut{Input}{Input}
\SetKwInOut{Output}{Output}
\SetKwInOut{Leaning}{Learning}
\SetKwInOut{Test}{Test}
\SetKwInOut{Initialize}{Initialize}
\SetKwFunction{FMain}{DULQA}
 \Input{$\tau,\eta_0,\gamma_0,N_{\mathrm{epoch}}$}
 \Initialize{$\{\eta(t)\}_{t = 0}^{\tau}\leftarrow \eta_0,\{\gamma(t)\}_{t = 0}^{\tau}\leftarrow \gamma_0$}
 \For {$t = 1,\dots,\tau$}{
 \For{$n = 1,\dots,N_{\mathrm{epoch}}$}{
 \text{Generate dataset $\mathcal{D}$}\\
\For{$d = 1,\dots,|\mathcal{D}|$}{
$\hat{\bm{w}}^d\leftarrow$\\
\text{DULQA($\bm{J}^d,\bm{h}^d,t,\bm{w}^d(0),\{\eta(t')\}_{t' = 0}^{t},\{\gamma (t')\}_{t' = 0}^{t}$)}
}
\text{Compute loss function $L(\mathcal{D})$ }\\
\text{Update the parameters via backpropagation }
}
 
 }
   \SetKwProg{Fn}{Function}{:}{}
    \Fn{\FMain{$\bm{J},\bm{h},\tau,\bm{w}(0),\{\eta(t)\}_{t = 0}^{\tau},\{\gamma(t)\}_{t = 0}^{\tau}$}}{
       \For{$t = 0,\dots,\tau$}{
 $\bm{w}(t + 1)\leftarrow \bm{w}(t) - \eta(t)\nabla_{\bm{w}(t)}C(\bm{w}(t),\bm{J},\bm{h},s(t),\gamma(t))$ 
\\
$s(t)\leftarrow \frac{t}{\tau}$
 }    
}
 \textbf{return} $\bm{w}(\tau+1)$ \\

\end{algorithm}

\section{Experiments}
\label{sec4}
In this section, we present the numerical results for DULQA.
First, we explain the problem and parameter settings for DULQA. Next, we present the numerical results of DULQA for the training and test data.
\subsection{Problem and parameter settings}
\label{sec41}
We consider the Sherrington--Kirkpatrick model \cite{Sherrington_1975} as follows: {
$\hat{H}_0 = \sum_{i\neq j}J_{ij}\hat{\sigma}_i^z\hat{\sigma}_j^z/\sqrt{N}$}, where $J_{ij}$ is generated randomly from a Gaussian distribution with zero mean and unit variance.
We implement DULQA using PyTorch \cite{Paszke_2019}.
We set the system size to $N = 1000$, annealing time to $\tau = 20$
, number of epochs to $N_{\mathrm{epoch}} = 5000$, amount of data to $|\mathcal{D}| = 200$, initial {step size} to $\eta_0 = 0.1$, and initial scale parameter to $\gamma_0 = 1$.
{The scale parameter of $\bm{w}^d(0)$ is fixed to $f = 0.5$ used in the implementation of the original LQA \cite{LQA_implementation} since the qualitative behavior of DULQA does not change in the range of $ 0.1\leq f \leq 1$ from our preliminary experiments. }
Trainable parameters are updated using the Adam optimizer \cite{kingma_2017} with an initial {learning rate} of $10^{-3}$.

\subsection{Numerical results}
\label{sec42}
\subsubsection{Performance for training dataset}
\label{sec421}
\begin{figure}[t]
\centering
\includegraphics[width=74mm]{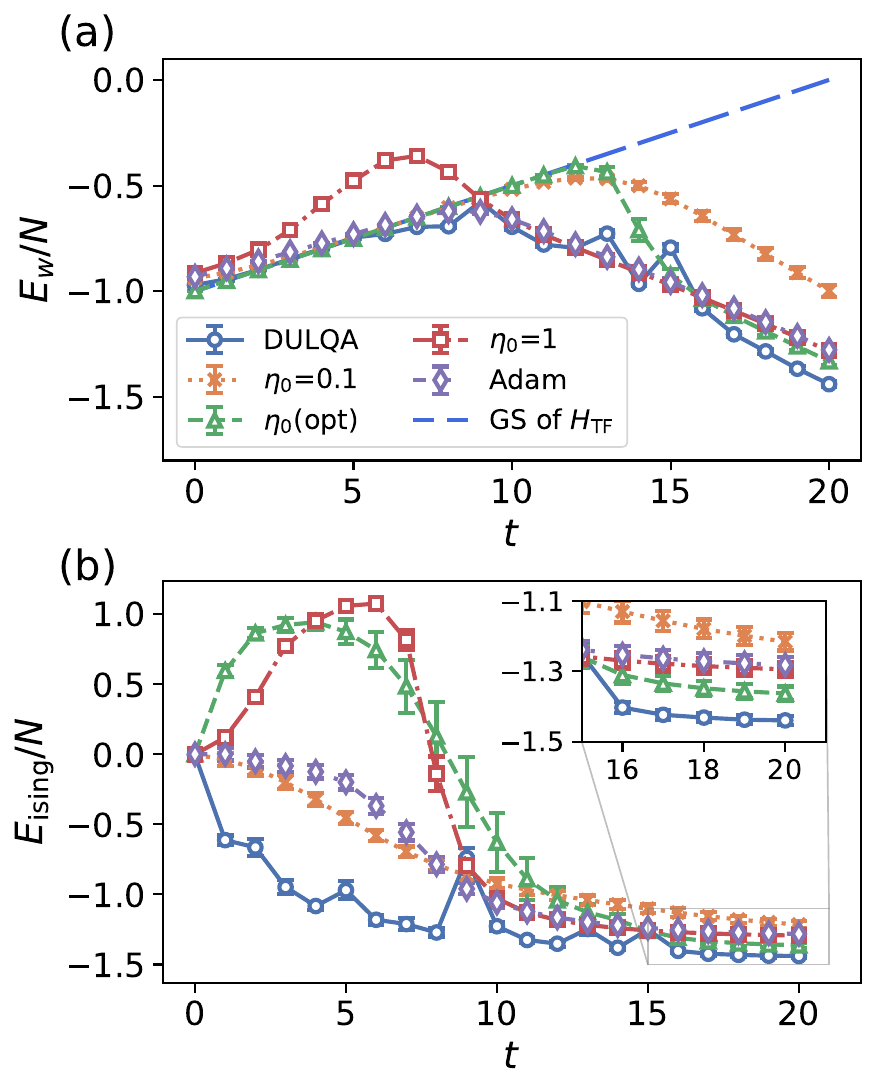}
\caption{Dependence of (a) $E_w/N$ and (b) $E_{\mathrm{ising}}/N$ on $t$ obtained by DULQA and LQA using GD with the fixed step size and Adam.
The straight line represents the ground state energy of the second term in Eq. \eqref{eq6}. {The inset shows the enlargement in the region $15\leq t\leq 20$. }
}
\label{fig:fig3}
\end{figure}

First, we consider the $1$-instance learning.
We fix problem $\bm{J}$ during training and verify whether the training of the parameters works. 
We compare the performance of DULQA with that of LQA using GD with a fixed step size $\eta_0$ and Adam.
Figure. \ref{fig:fig3} shows the dependence of $E_w /N$ and $E_{\mathrm{ising}}/N$ (see Sec. \ref{sec2}) on $t$.
The symbols are averaged over the data, and the error bars represent standard deviations (SDs).
All error bars in the figures in this subsection are calculated in this manner.
{The {trained} parameters of DULQA with 1-instance learning are shown in Appendix \ref{app1}.}
For LQA, we set $\gamma = 1$.
The optimal step size of GD is computed using Optuna \cite{Akiba_2019} and is written as $\eta_{0}$(opt).
The initial step size of Adam is optimized in the same manner.
Figure. \ref{fig:fig3} (a) shows that the performance of DULQA is better than that of LQA using GD and Adam at $t = 20$.
The dashed straight line denotes the instantaneous ground state energy of the transverse field term $\hat{H}_{\mathrm{TF}}$.
Except for the result of GD with $\eta_0 = 1$, the results of LQA and DULQA remain around the line in the short time region.
This behavior is similar to the adiabatic time evolution of the vanilla QA.
In addition, DULQA escapes the minima of the transverse field term faster than LQA and improves the approximation performance.
This is similar to a previous study \cite{Bittel_2021} where the momentum method leads to a fast escape from the ground state of $\hat{H}_{\mathrm{TF}}$ and an enhancement of the performance, possibly because it emulates QA dynamics more accurately.
This suggests that the learning step sizes in DULQA determine (sub)optimal QA-like dynamics required to minimize the energy at $t = \tau$.

As shown in Fig. \ref{fig:fig3} (b), LQA initially stays at approximately zero for or increases the $E_{\mathrm{ising}}/N$.
In contrast, DULQA gradually decreases $E_{\mathrm{ising}}/N$ from the early annealing stage.
This indicates that training proceeds in the direction of minimizing $E_{\mathrm{ising}}$ using the DU scheme, resulting in a fast escape from the ground state of $\hat{H}_{\mathrm{TF}}$. 

\begin{figure}[t]
\centering
\includegraphics[width=70mm]{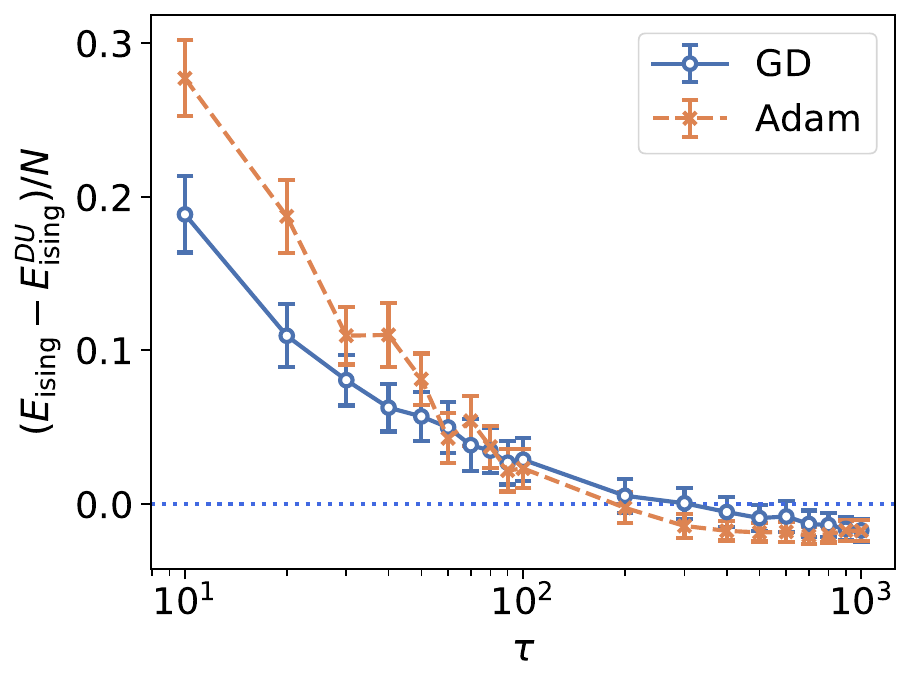}
\caption{Performance difference of $E_{\mathrm{ising}}/N$ obtained by LQA and DULQA.
The horizontal axis denotes the annealing time.
}
\label{fig:fig4}
\end{figure}
Next, we demonstrate how {long} annealing times LQA needs to exceed the performance of DULQA in Fig. \ref{fig:fig4}.
 We perform LQA using GD and Adam under $200$ different initial conditions for a different $\tau$.
 During the simulation, we fix the instance utilized in Fig. \ref{fig:fig3}. 
 Optuna tunes the parameters of GD and Adam for each $\tau$.
 We compute the difference of $E_{\mathrm{ising}}/N$ obtained by LQA and the minimum value of $E_{\mathrm{ising}}/N$ attained by DULQA at {$t = 20$} used in Fig. \ref{fig:fig3}. 
 The minimum value of $E_{\mathrm{ising}}/N$ is computed from $200$ independent trials.
 From Fig. \ref{fig:fig4}, LQA with GD and Adam requires more annealing time than DULQA to reach its performance.
DULQA can attain lower energy approximately ten times faster than LQA using GD with a fixed step size and Adam.
For a fixed instance, the DU scheme improves the performance of LQA.

\subsubsection{Performance for test dataset}\label{sec422}
It is desirable that the {trained} DULQA exhibits performance improvement not only for a problem in the training dataset but also for other problems. 
In this subsection, we investigate the latter performance and generalization performance of DULQA.

\begin{figure}[t]
\centering
\includegraphics[width=70mm]{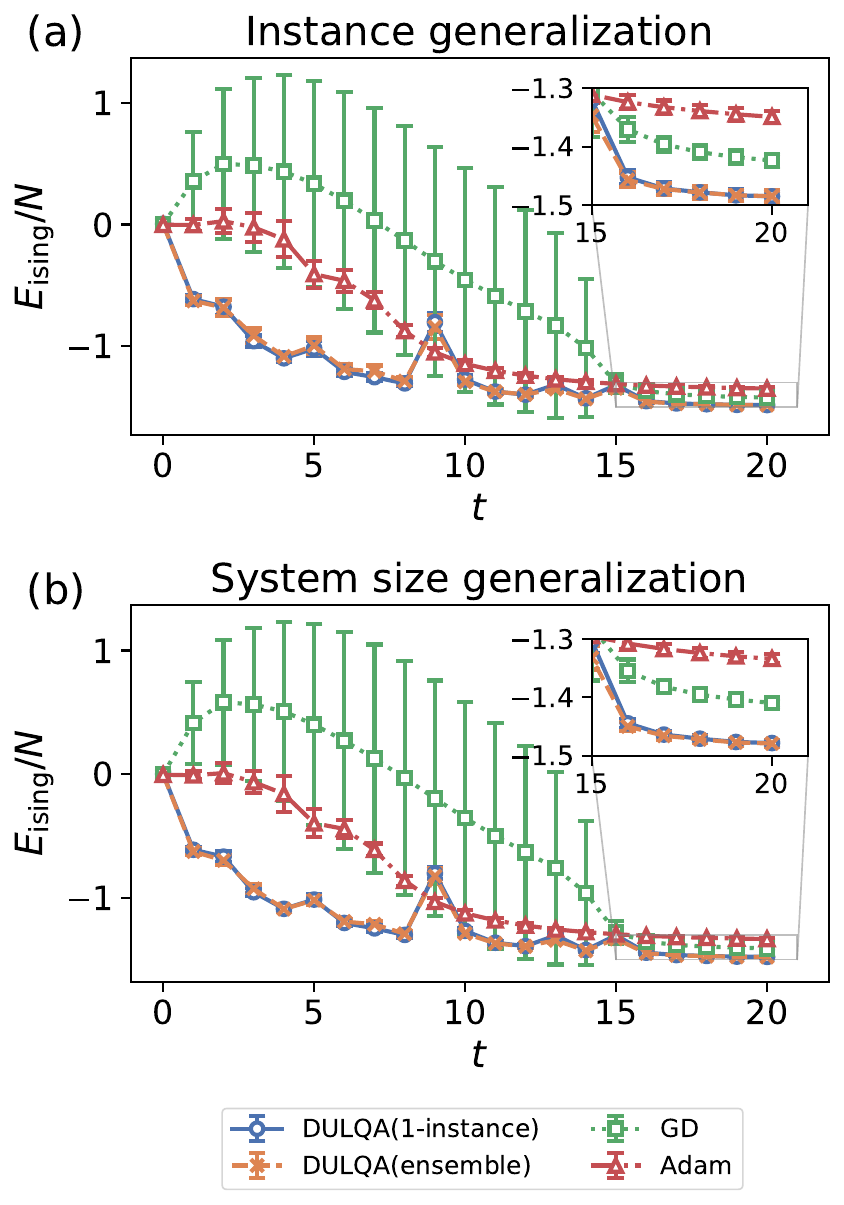}

\caption{Generalization performance of DULQA 
(a): for different instances with $N = 1000$ and (b): for different instances with $N = 2000$.
The vertical axis represents $E_{\mathrm{ising}}/N$. The horizontal axis denotes the time $t$. The parameters of the "DULQA(1-instance)" are the same as used in Fig. \ref{fig:fig3}. {Each inset enlarges the region $15\leq t\leq 20$. }
 }
\label{fig:fig5}
\end{figure}

Here, we evaluate $ E_{\mathrm{ising}}/N$ for an unknown instance and different system sizes.
In Fig. \ref{fig:fig5} (a), we fix the system size $N = 1000$ and evaluate the performance of DULQA for test instances.
Each symbol denotes the average of over $100$ test instances, and the error bar represents the SD.
In Fig.~\ref{fig:fig5}, ``DULQA(1-instance)" represents results obtained by the {trained} parameters in the experiment in Fig. \ref{fig:fig3}. 
We add results of ``DULQA (ensemble)" where parameters of DULQA are learned by ensemble learning whose datasets contain multiple realizations of $\{J_{ij}\}$.
{The trained parameters of DULQA with ensemble learning are also exhibited in Appendix \ref{app1}.}
For comparison, the parameters of LQA with GD or Adam are optimized using Optuna for each test instance to improve the performance.
Figure. \ref{fig:fig5} (a) exhibits that the performance of the ``DULQA(1-instance)" and ``DULQA(ensemble)" outperform that of LQA using GD and Adam. 
Although the error bars for DULQA and LQA using Adam are small, those for LQA using GD become large in the intermediate annealing region.
The performance and trajectory of LQA utilizing GD largely depend on the instance.
{The parameters of DULQA and LQA using Adam changes at each time step. Since the time-dependent controls of parameters can reduce fluctuations of the optimization trajectory for each instance in the intermediate annealing region, they lead to the stable behaviors of $ E_{\mathrm{ising}}/N$. 
Note that {higher energy of LQA with GD in the intermediate region does not imply poorer results at convergence. 
This is because a step size of LQA with GD is optimized for each problem to minimize the final energy, neglecting the energy in the intermediate region.}}
The difference in performance between DULQA with 1-instance learning and ensemble learning is small.
In particular, unlike the original LQA, DULQA uses {trained parameters that are irrelevant} to test instances. This demonstrates the remarkable generalization performance of DULQA.

\begin{figure}[t]
\centering
\includegraphics[width=70mm]{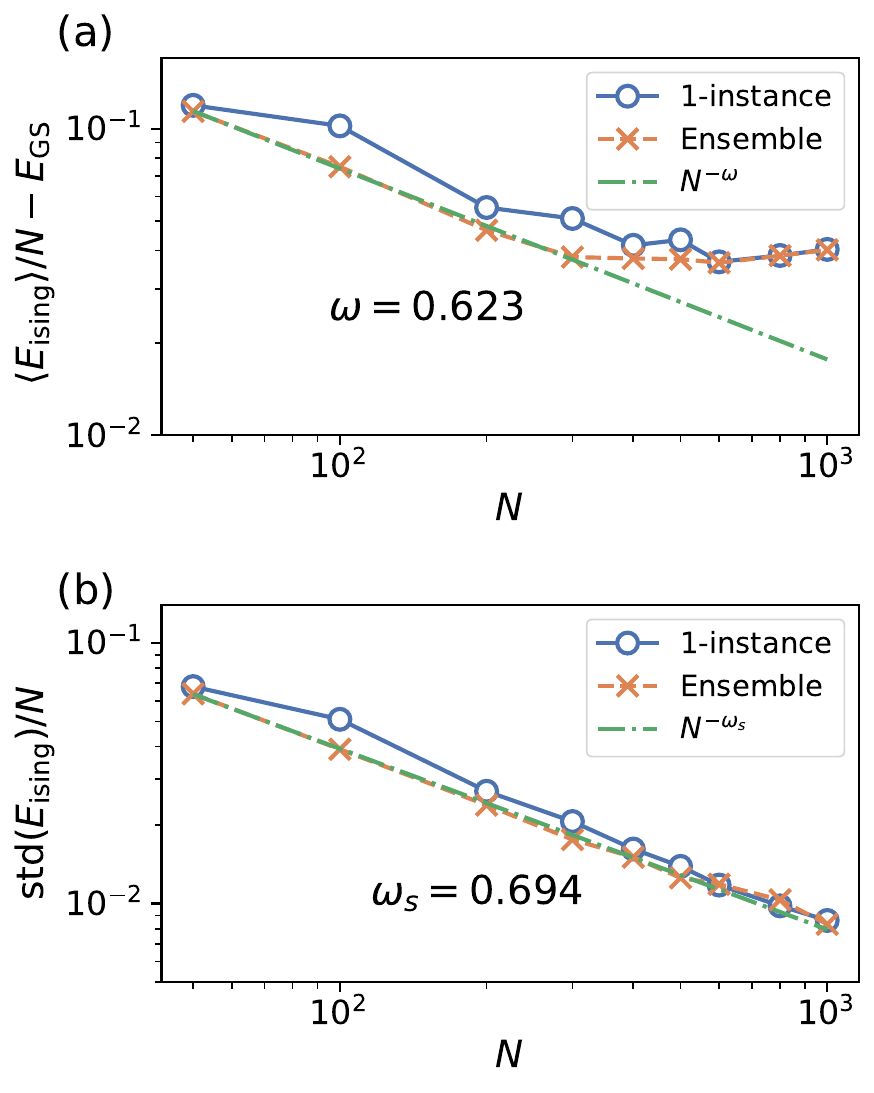}
\caption{Dependence of the statistic of $E_{\mathrm{ising}}/N$ obtained by DULQA with 1-instance and ensemble learning on the system size $N$.
The vertical axis denotes the statistics of $E_{\mathrm{ising}}/N$ : (a) residual energy and (b) SD.
The dash-dotted line represents the fitting line for the data of ensemble learning for the residual energy in $50\leq N \leq300$ and the SD in {$50\leq N\leq1000$.}
}
\label{fig:fig6}
\end{figure}

Next, we demonstrate the performance of DULQA for test instances of different sizes $N = 2000$ in Fig. \ref{fig:fig5} (b).
The performance of DULQA surpasses that of LQA with GD and Adam.
Therefore, the trained parameters of DULQA are applicable to different instances or instances of larger system sizes without losing performance, which is desirable for practical applications.
Note that the parameters of LQA are optimized for each instance, leading to additional preprocessing time for parameter optimization during execution. 
In contrast, the DU scheme offers a learned LQA applicable to an ensemble of problems without parameter optimization once the training process is completed.

Finally, we compare 1-instance learning with ensemble learning for DULQA with various $N$ values to study the difference and finite-size effects of the two learning strategies.
{We show the residual energy between the average of $E_{\mathrm{ising}}/N$ and the ground state energy $E_{\mathrm{GS}}\simeq -1.526$ \cite{Parisi_1979} in the thermodynamic limit \footnote{In our paper, the definition of the Hamiltonian is different. We take summation over $i\neq j$. In the original formulation, the summation is taken over $i < j$. Therefore, we utilize the doubled ground state energy shown in \cite{Parisi_1979}.} and the SD of $E_{\mathrm{ising}}/N$ in Fig. \ref{fig:fig6}.
These statistics are computed over $500$ test instances.
For each instance, DULQA was performed 200 times under different initial conditions. The minimum values of $E_{\mathrm{ising}}/N$ are used to compute the statistics. Note that the data for $N = 1000$ differ from those in Fig. \ref{fig:fig5} (a). 
Figure. \ref{fig:fig6} (a) shows that ensemble learning leads to a higher average performance by incorporating data fluctuations because the small instance has a high dispersion. The dash-dotted line denotes the fitting line of the ensemble learning results.
The residual energy decays polynomially with respect to $N$ for $50\leq N\leq 300$. 
The obtained exponent $\omega = 0.623$ is similar to the value $\omega = 2/3$ \cite{Palassini_2008} and larger than the value $\omega = 0.61$ obtained from the mean-field approximation optimization algorithm (MAOA) \cite{Spieldenner_2023}.
The saturated residual energy for $N > 300$ reflects the difficulty of training DULQA for a larger $N$, and we may need more data or epochs.
For a larger $N$, the difference in residual energy between 1-instance learning and ensemble learning decreases.
This {behavior is due} to the self-averaging property of the SK model.
As the average performance is replaced by the performance of an individual instance for a large $N$, no significant difference in the statistics of the residual energy between 1-instance learning and ensemble learning exists.
{
As explained in Sec. \ref{sec22} and \ref{sec3}, the DU scheme learns the statistical properties of problem instances.
Therefore, it is noted that this behavior can be seen not only in the SK model but also in other problem instances with the self-averaging property.
{If test instances are generated from a problem set different from the training dataset, a model mismatch occurs, potentially leading to unexpectedly poor performance in DU-based algorithms.}
}

{Figure. \ref{fig:fig6} (b) shows that the difference in the SD between 1-instance and ensemble learning is slight owing to the self-averaging property.
The SD of $E_{\mathrm{ising}}$ decays polynomially with respect to $N$ and converges to zero in the $N\rightarrow \infty$ limit. 
{The exponent of the SD $\omega_s = 0.694$ is larger than that of the MAOA $\omega_s = 0.68$ \cite{Spieldenner_2023}.   
This indicates that DULQA has the potential to surpass the MAOA. 
Note that we consider whether the exponent is universal to other approximated algorithms.}

\section{Conclusion}
\label{sec5}
In this study, we proposed DULQA based on the DU scheme and applied it to the SK model.
DULQA learns the parameters {in the optimization process of LQA} from the training dataset, whereas the performance of the original LQA depends on the parameters.
We also applied two learning strategies: 1-instance and ensemble learning.
The parameters obtained by DULQA led to faster convergence and better performance than those obtained by LQA. 
Numerical studies have shown that the {trained} parameters can be successfully generalized to different instances with the same or larger system size.
The performance of DULQA surpassed that of LQA with GD or Adam, with the parameters optimized by Optuna.
For small instances, the performance of DULQA with ensemble learning was better than that with 1-instance learning.
The difference in performance between 1-instance and ensemble learning was fairly small for large-sized instances owing to the self-averaging property of the SK model.
For a fixed annealing time, the parameters that returned good solutions for some instances also yielded similar performance for unknown instances. 
This behavior can be observed in the QAOA \cite{Fernando_2018, Farhi2022,takabe_2024}. 
In addition, the system size dependence of the residual energy and SD reproduced similar exponents obtained in a previous study \cite{Palassini_2008, Spieldenner_2023,10345665}.
For a small $N$, the residual energy was polynomially scaled with $N$.
{The exponents of the residual energy and SD of the energy obtained by DULQA were slightly larger than those obtained by the MAOA.}
These results demonstrate the potential of DULQA.
However, for a large $N$, the residual energy was saturated.
Training the parameters for a large $N$ is an important research direction.

{It is worth mentioning the connection between DULQA and QAOA. 
QAOA learns the variational parameters by minimizing the energy expectations.
DULQA trains a time-dependent step size and coefficient by minimizing the loss function.
As stated in \cite{Zhou_2020}, QAOA utilizes a diabatic transition instead of adiabatic evolution.
DULQA emulated the adiabatic-like QA dynamics. 
The two algorithms are similar in that they utilize optimized parameters; however, their operating principles and acceleration mechanisms are different.
A detailed investigation of the physical interpretation of the acceleration mechanism is required. 
}
}

We focused on learning $\eta(t)$ and $\gamma(t)$.
Rather than learning $\gamma(t)$, we can learn a QAOA-like scheduling function parameterized by $s(t)\gamma \rightarrow A(t)$ and $1 - s(t)\rightarrow B(t)$.
As we increase the number of parameters, the performance of DULQA is enhanced.
In this study, the annealing time is fixed.
A long annealing time increases the degrees of freedom and enhances the performance of DULQA, similar to QA and QAOA.
However, increasing the number of parameters makes the optimization process difficult because a deep structure often results in gradient vanishing and learning instability. {The initialization strategy for utilizing discrete sine and cosine transformations presented in \cite{Zhou_2020} may be a promising approach.
Heuristic strategies are an exciting topic for future studies.}
In this study, we considered the SK model as the first benchmark for DULQA. 
DULQA can be applied to other combinatorial optimization problems.
As in the case of vanilla QA, the potential impact of non-stoquastic terms \cite{seki_2012} or reverse annealing \cite{Ohkuwa_2018, Yamashiro2019, Arai_2021} on the performance of DULQA is an intriguing and engaging topic that warrants further exploration.

\section*{Acknowledgment}
This study was partially supported by JSPS KAKENHI with grant numbers 22H00514 and 22K17964. 
\appendix 
{
\section{Trained parameters of DULQA}
\label{app1}
We show the time dependence of the trained parameters of DULQA obtained by two different strategies in Fig. \ref{fig:fig7} {when $N=1000$. The qualitative difference in the behaviors between 1-instance learning and ensemble learning does not exist, leading to a similar performance for test datasets.}
The zig-zag behavior of $\eta(t)$ and $\gamma(t)$ can be found just like the DU-GD algorithm \cite{takabe_2021}.
In the early annealing stage where the transverse field term is dominated, the learned $\eta(t)$ takes a smaller value.
It is interpreted that the small step size reduces the fluctuation of the optimization trajectory.
Except for at the end of the annealing stage, inverse proportional behaviors of $\eta(t)$ and $\gamma(t)$ can be seen.
At the end of the annealing stage, both $\eta(t)$ and $\gamma(t)$ increases. 
The theoretical understanding and interpretation of trained parameters are needed and remain an open question except for the case of the DU-GD algorithm \cite{takabe_2021}.}
\begin{figure*}[t]
\centering
\includegraphics[width=150mm]{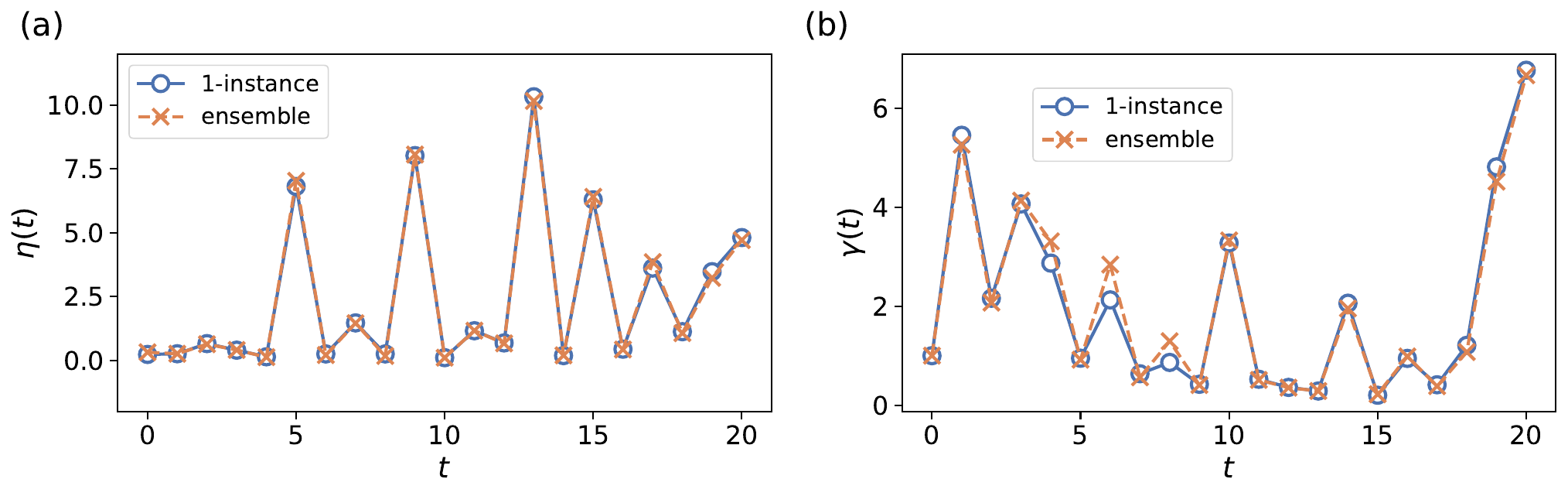}
\caption{Time dependence of the trained parameters of DULQA obtained by two different strategies (a): $\eta(t)$ and (b): $\gamma(t)$. }

\label{fig:fig7}
\end{figure*}
\bibliography{main.bib}
\end{document}